\documentclass{article}

\usepackage{PRIMEarxiv}
\usepackage{amsmath}
\usepackage[utf8]{inputenc} 
\usepackage[T1]{fontenc}    
\usepackage{hyperref}       
\usepackage{url}            
\usepackage{booktabs}       
\usepackage{amsfonts}       
\usepackage{nicefrac}       
\usepackage{microtype}      
\usepackage{lipsum}
\usepackage{placeins}
\usepackage{fancyhdr}       
\usepackage{graphicx}       
\graphicspath{{figures/}}     
\usepackage{natbib}

\title{Distilling interpretable causal trees from causal forests
}

\author{
  Patrick Rehill\\
  POLIS: The Centre for Social Policy and Research \\
  Australian National University \\
  Canberra, Australia\\
  patrick.rehill@anu.edu.au \\
}

\begin{document}
\maketitle

\begin{abstract}
Machine learning methods for estimating treatment effect heterogeneity promise greater flexibility than existing methods that test a few pre-specified hypotheses. However, one problem these methods can have is that it can be challenging to extract insights from complicated machine learning models. A high-dimensional distribution of conditional average treatment effects may give accurate, individual-level estimates, but it can be hard to understand the underlying patterns; hard to know what the implications of the analysis are. This paper proposes the Distilled Causal Tree, a method for distilling a single, interpretable causal tree from a causal forest. This compares well to existing methods of extracting a single tree, particularly in noisy data or high-dimensional data where there are many correlated features. Here it even outperforms the base causal forest in most simulations. Its estimates are doubly robust and asymptotically normal just as those of the causal forest are.
\end{abstract}


\section{Introduction}
The development of causal machine learning methods that estimate treatment effect heterogeneity fundamentally changes the kinds of research questions that can be answered. Now drivers of heterogeneity can be explored in a data-driven way without needing to worry about multiple testing \citep{shiba_harnessing_2024} and it is possible to obtain accurate, individual-level Conditional Average Treatment Effect (CATE) estimates. Individual-level estimates are very useful in some applications such as in personalised medicine or marketing, where the goal is to target specific treatments to specific people. However, where the goal is to make high-level decisions about an intervention informed by differences in treatment effect --- such as in policy evaluation where individual-level targeting is generally not possible or desirable --- it can be hard to know how to extract useful insights from a model. For example, in the literature using the causal forest --- arguably the most popular causal machine learning method for this kind of application --- there does not seem to be an agreed-upon way to resolve this problem \citep{rehill_transparency_2023}. The methods used generally fall back on exploring effects across theoretically important variables or those variables deemed important by a variable importance approach that has real flaws when data are noisy and high-dimensional \citep{benard_variable_2023}. These approaches also generally ignore interaction effects between variables as they are interested only in bivariate distributions of CATE against a given variable \citep{shiba_harnessing_2024, rehill_2024_how)}. One approach used in a handful of papers is to extract a single interpretable causal tree \citep{rehill_2024_how)}. This approach, pulled from the interpretable machine learning literature \citep{rudin_stop_2019} can be used to summarise the distribution of CATEs in a single tree structure given understandable clusters of effect (leaves) and accounting for interactions between features (the tree structure). The problem is though that these trees are poor learners, they are often unstable and they are only fit on a subset of all features giving an incomplete picture of heterogeneity.

This paper proposes an approach to improve the ability of a single tree to fit the underlying CATE distribution using all available features and yielding doubly robust estimates for each leaf. I term this approach a Distilled Causal Tree (DCT). Rather than extracting a single tree from an ensemble as many current approaches do \citep{rehill_2024_how)}, the DCT uses an approach called knowledge distillation (KD) \citep{frosst_distilling_2017, dao_knowledge_2021} to fit a better approximation of the CATE distribution. The key insight of KD is that rather than fitting an interpretable model like a decision tree directly on raw data, one can get a more powerful model by fitting a complex, black-box learner (called a teacher) on the data first. Then the interpretable model is fit not on the raw data (as a basic model would be), but on the predictions from the teacher (we call this the distilled or student model). In many contexts, the teacher's predictions actually hold more information about the underlying data-generating process (DGP) than the raw data does and so this distilled model outperforms the basic model. In the case of KD in a classification problem, this better knowledge of the underlying DGP is data that provides an estimate of Bayes class probabilities (the actual probability of being in a given class for an observation) \citep{dao_knowledge_2021}. In the case of causal inference, this underlying distribution is ground-truth causal effects.

Unfortunately, not all teachers are amenable to distillation. Luckily for us, the kinds of attributes that make for a good CATE learner also make for a good teacher as I will explain in Sections \ref{sec:3} and \ref{sec:dct}. This means that we can be reasonably confident that a distilled tree will fit the CATE distribution better than a non-distilled tree.

Another insight discussed in Section \ref{sec:evtrees} is that given KD turns the fitting of a causal tree into a standard regression problem (predicting the predictions of the teacher) we can apply a range of different algorithms to that problem, not just being limited to the standard causal tree \citep{athey_recursive_2016}. These new algorithms can provide a better model. To put it another way, other approaches involve extracting a single good causal tree from the forest, the DCT gives us the flexibility to separate out the teacher fitting approach (where we need to fit many trees quickly to estimate a large ensemble \citep{athey_generalized_2018, breiman_random_2001}) from the fitting of the single interpretable tree (where we can afford to invest more computation in improving the tree's fit). Standard tree fitting approaches are greedy, meaning that they split recursively, never revisiting previous splits. This paper proposes using an optimal tree fitting algorithm (I focus on an evolutionary algorithm here) which can be computationally expensive, but can produce better trees because it can optimise all splits at the same time. These two enhancements taken together can boost the ability of a single tree to explain variation in treatment effects with a model that is still interpretable.

In Section \ref{sec:sim} I show that in a simulation study not only does the distilled tree generally outperform all the tree extraction approaches, it can also outperform a full causal forest in high-dimensional datasets with a low signal-to-noise ratio. Section \ref{sec:application} applies the method to an application taking data from a field experiment on reducing transphobia with door-to-door canvassing. Code implementing the DCT, the experiments and the application can be found at \url{https://github.com/pbrehill/DistilledCausalTree}.

\section{Problem set-up}
\subsection{Potential outcomes and causal inference}
In a causal inference problem we aim to estimate a treatment effect $\tau$ rather than predict an observed value. In the Potential Outcomes framework \citep{imbens_causal_2015}, for a binary treatment, this means we are estimating the difference between two potential outcomes, that in which an individual is treated and that in which the individual is not; $\tau = Y(1)-Y(0)$. This makes the problem of estimation fundamentally different because the ground-truth values for the quantity we wish to estimate are never observable. This is called the Fundamental Problem of Causal Inference \citep{holland_statistics_1986}. This has been particularly problematic in the machine learning context. The theoretical properties of for example a linear regression model may allow for unbiased estimation of treatment effects just by fitting on observed outcomes, but unfortunately, most complicated machine learning models lack such properties, making the naive use of machine learning models for causal inference ill-advised. Statistical breakthroughs like the work of  \citet{semenova_debiased_2021}, \citet{kunzel_metalearners_2019}, \citet{athey_recursive_2016}, \citet{wager_estimation_2018} and others have all created ways to get around these limitations, yielding estimators with better statistical properties than naive approaches that just aim to fit outcomes and infer causal effects from there \citep{kunzel_metalearners_2019}.

There are broadly two kinds of tasks we might be interested in using causal ML to solve, the problem of modelling out confounding, and the problem of explaining treatment effect heterogeneity. In the first case, our focus is on modelling out selection effects in order to estimate some kind of causal quantity that would be trivial to estimate in unconfounded data, for example an Average Treatment Effect (ATE) $ATE=\mathbb{E}[\tau]$ or an Average Treatment Effect on the Treated (ATT) $ATT=\mathbb{E}[\tau|W=1]$. Here machine learning approaches like double machine learning \citep{chernozhukov_doubledebiased_2018} are useful.

In the second kind of application we are interested in estimating not a single causal quantity, but a distribution of effects so machine learning tools become useful for mapping out that distribution, for example a Conditional Average Treatment Effect (CATE), $CATE=\mathbb{E}[\tau|X]$. These two approaches are also commonly combined such as where a causal forest is used to estimate a CATE in confounded data by employing local centering and a doubly robust estimator \citep{athey_generalized_2018}. This uses nuisance models estimating outcomes and treatment to debias (in case of doubly robust estimation) or increase the efficiency of (in the case of local centering) the CATE estimator. For the purposes of this paper though, we will assume unconfounded data such as from an experiment. In practice this approach can be applied to cases where we meet the Independence Assumption in other ways, but this is orthogonal to how we then estimate out CATEs.

\subsection{Knowledge distillation}
KD is a field that aims to fit an interpretable model that is as powerful as possible under these interpretability constraints. It does this by not fitting an interpretable model on raw outcomes, but instead by modelling the predictions of a black-box 'teacher' \citep{frosst_distilling_2017} with an interpretable model creating a 'student' model. For example, a decision tree is a common student model \citep{frosst_distilling_2017, domingos_knowledge_1997, dao_knowledge_2021, rudin_stop_2019}.

Fitting on the outcomes of the teacher rather than raw data can substantially boost the performance of a student learner over how it would perform if just fit on the raw data. In fact in some cases, an interpretable student can perform as well or better than the teacher model \citep{dao_knowledge_2021}. This is a stunning finding made even stranger by the fact that for a very long time, we lacked good theory on why some situations were amenable to knowledge distillation and others were not. The key paper to attempt to lay out theory on this is \citet{dao_knowledge_2021}. That paper underpins the theoretical approach to KD taken in this paper. While it uses examples in from classification, knowledge distillation is also applicable to regression contexts (like learning CATEs) for example in \citet{lee_teacher_2018} and \citet{wang_model_2017}. All that should be needed here is to use a regression distillation loss rather than a classifcation distillation loss (the actual characteristics of Dao et al.'s enhanced KD are agnostic to the specific loss function).

Dao et al. explain knowledge distillation as a semi-parametric inference problem. The idea here is that the teacher acts as a plug-in estimator for underlying ground-truth information that is not available in the raw data. For example, in a classification problem we only have access to realised labels in raw data (which are binary) but a teacher's predictions provide a probability for each class. This is a nuisance estimate of the underlying Bayes class probabilities.

This is a good theory of why knowledge distillation works, but why does it work well in some contexts and not others? Dao et al. explain that there are properties of the teacher that can affect performance. Essentially the teacher should not underfit or overfit the underlying distribution. If it does, it will not perform well as a nuisance model. However, viewing the problem through this lens, there are ways to alter the problem of learning a teacher that will improve its functioning as a teacher (even if this is at the cost of actual accuracy of the teacher). Dao et al. propose two adjustments. The first is cross-fitting as proposed by \citet{chernozhukov_doubledebiased_2018}. Here we essentially fit models inside different folds and then predict results out-of-fold. This combats overfitting of the teacher. The second adjustment is bias correction to correct for underfitting. They argue the plug-in KD loss is a zeroth order Taylor approximation of the ideal loss that is the loss comparing estimates to Bayes probabilities (or the analogous ground-truth treatment effects). They suggest correcting for bias by increasing the order of the Taylor approximation using a noisy but unbiased estimate of the Bayes class probabilities --- the raw labels. The actual approach is slightly more complicated and a full explanation can be found in that paper. For our purposes though, the DCT proposal will actually use a different method entirely for removing bias --- guarantees of asymptotic normality --- and so the actual method proposed in Dao et al. is not particularly important.

The use of cross-fitting and bias correction make up what Dao et al. call 'enhanced KD' which generally outperforms 'vanilla KD' which lacks these components. Those familiar with the causal machine learning literature are likely already seeing some similarities between enhanced KD and causal machine learning estimators. The way we learn CATES in a method like the causal forest is also by semi-parametric inference (in this case using two nuisance models --- one for expected outcome, one for expected treatment --- to construct an estimate of ground-truth causal effects). In addition, cross-fitting will also be familiar, the paper on the concept cited by Dao et al. is the seminal \citet{chernozhukov_doubledebiased_2018} which proposed double machine learning and the causal forest borrows the concept though operationalises it as out-of-bag predictions in the forest ensemble rather than it k-fold cross-fitting \citep{shiba_harnessing_2024, wager_estimation_2018}.

In the following section, I will argue that these similarities in language are not merely coincidences. The theoretical guarantees needed for good causal inference also make causal forest models well suited to being teachers in a KD problem. In a sense, when distilling from a causal forest we get enhanced KD for free.

\section{A knowledge distillation perspective on causal inference} \label{sec:3}
\subsection{Applying KD to the causal forest}
When it comes to distilling out a causal model, there are a couple of changes from the standard knowledge distillation problem that Dao et al. tackle. Many of these differences, I will address in the next section, but it is worth discussing how one understands the two KD enhancements from Dao et al. in a causal machine learning context first.

To start with cross-fitting, we see a clear commonality between the causal forest and enhanced KD. As previously discussed, the causal forest uses cross-fitting to address overfitting. Both in its nuisance models and the actual heterogeneity model itself. The causal forest uses out-of-bag estimation which functions exactly as cross-fitting does (though is computationally simpler for a random forest-like ensemble) \citep{wager_estimation_2018, shiba_harnessing_2024}. This means that we already have cross-fitting built in.

But what to do about bias correction? This is more of a challenge. The problem is we cannot simply plug in realised outcomes as Dao et al. do because we do not have realised causal effects due to the Fundamental Problem of Causal Inference. One solution might be to deploy the pseudo-outcomes from R-Learner \citep{nie_quasi-oracle_2020} here which are estimates of the underlying effects we are trying to recover just as outcomes are in Dao et al.'s KD. However, in practice these are likely to be extremely noisy and I am not aware of any proof of the unbiasedness of the pseudo-outcomes in the way raw class labels are unbiased in expectation. A better approach then is to rely on the asymptotic unbiasedness that is built into the causal forest.

Assuming unconfoundedness, the causal forest is already asymptotically unbiased thanks to honest estimation \citep{athey_recursive_2016, athey_generalized_2018}. While this appears to be a less general set of conditions than the Taylor approximation-based bias correction proposed by Dao et al., this is not really the case. We need a sufficiently large sample to rely on asymptotic properties and we need to assume causal identification regardless of whether we are distilling or not in order to make valid causal inferences with a causal forest. In a sense then, we also get a version of bias correction 'for free' when distilling causal forests.

As the model already incorporates cross-fitting and bias correction, there is no need for modifications to the loss function used in the DCT as there are for other distillation problems in Dao et al.. We can simply use mean squared error of student predictions against teacher predictions for DCT distillation.

\subsection{How KD may improve performance where there are many noisy and correlated features} \label{sec:correlation}

One problem that researchers face when trying to fit a causal forest on real-world data is the correlation of features in a high-dimensional dataset. This is a problem both for the extraction of insights and for the accuracy of models.

On extracting insights from a causal forest, we rely on the ability to pull out measures of important variables to do this. This could be actual variable importance metrics, or in the case of a single-tree approach, this could be about pulling out trees that split on variables that tell us something about the underlying CATE distribution. Where features are correlated, it can be difficult to separate out the effects of one variable from another, resulting in the standard causal forest variable importance approach 'diluting' the importance of variables with many other correlated variables in the dataset \citep{benard_variable_2023}. So we might miss important underlying constructs simply because we have too many measures of that construct in the dataset.

Another problem is around accuracy of the learner. Random forests and therefore causal forests rely on being able to sample different variables for different trees in order to improve their power. This is the key insight that separates a random forest from a simple ensemble of bagged trees \citep{breiman_random_2001, breiman_bagging_1996}. If there are many correlated measures of one construct in a dataset, it will not be possible to sample that construct out of the variables used to fit a tree resulting in something with the (generally lower) accuracy of an ensemble of bagged trees.

Of course while sampling a construct into too many trees can be a problem, there is also the accuracy problem of having too many noisy variables in a dataset meaning that key constructs are not representated in enough trees to accurately fit the underlying DGP.

Distillation may be helpful with both these problems. This is because per Dao et al.'s conceptualisation of KD, the process when done right essentially de-noises outcomes. The single tree can better ignore noisy features because they are less likely to correlate with a de-noised outcome. In addition, because it does not have to worry about working from a sample of variables, it can identify the variables that are the best predictors of these de-noised effects better than any single tree in a random forest could. This could yield a distilled tree that is better able to pick out the most important features yielding an interpretable and accurate model. This proposition is tested in the simulations in Section \ref{sec:sim}.

\section{The Distilled Causal Tree} \label{sec:dct}
Knowing we get Dao et al.'s enhanced KD out-of-the-box then, I propose a relatively simple procedure for fitting a DCT.

\textbf{Fit a teacher}

The first step in this process is to fit a teacher model. This could be any CATE learner though in this case we assume a causal forest. As already mentioned, the causal forest has the properties that allow for enhanced KD with no modifications. Just as if the causal forest itself were the learner of interest, it is important to grow a forest that drives excess error (error caused by too small an ensemble) down close to zero in order to get a well-performing teacher \citep{athey_generalized_2018}.

\textbf{Distill a single tree}

We can now fit the distilled tree based on the predictions from the teacher. As we want an honest tree, we must still only fit this on half of the sample, but this is otherwise a relatively standard predictive tree estimation problem. Rather than using the standard causal forest splitting function of Estimated Expected Mean Squared Error \citep{athey_recursive_2016}, we simply use the Mean Squared Error to approximate the treatment effect estimates of the teacher. To fit the teacher model $\hat{\tau}_T$ with an interpretable student $\hat{\tau}_S$ we minimise $$\text{MSE}= \frac{1}{n} \sum_{i=1}^{n} (\hat{\tau}_{T}(X_i) - \hat{\tau}_{S}(X_i))^2.
$$We fit this loss function using an arbitrary algorithm regression tree learner.

While hyperaparameter tuning is possible here, the guiding factor should be interpretability over maximising fit. Even if a large increase in the depth of a tree improves out-of-sample fit, it will dramatically hurt the interpretability of the model and at a certain point, the tree becomes almost as uninterpretable as the teacher. For this reason we set a depth limit for all learners here but allow them to fit shallower trees if the default criteria for further splitting are not met.

\textbf{Estimate out CATEs}

The final step is to make CATE estimates. Following \citet{athey_generalized_2018} we do not actually estimate out predictions from the distilled model but instead treat it as an adaptive kernel for a doubly robust estimator. This means taking the estimation sample of the distilled tree and re-predicting doubly robust estimates for each leaf. As the there is only a single tree in this case though, we are simply making a doubly robust estimate within the leaf using held-out data. To put it another way, it is like a causal forest estimator where all kernel weights are either zero or one. In this case we use Augmented Inverse Probability Weighting.

$$\hat{\tau}(x) = \frac{1}{n} \sum_{i=1}^n K_h(x, X_i) \left( \frac{W_i - \hat{e}(X_i)}{\hat{e}(X_i)(1 - \hat{e}(X_i))} (Y_i - \hat{\mu}_{W_i}(X_i)) + \hat{\mu}_1(X_i) - \hat{\mu}_0(X_i) \right), \quad K_h(x,X) \in \{0,1\}
$$

We can then estimate out standard errors by taking the tree structure as fixed (because due to honesty, the estimates are independent of the tree structure) and taking bootstrap estimates within each leaf. This is similar to the approach to standard error estimation in the \textit{grf} forest, though it relies on bootstrapping rather than jackknifing \citep{wager_estimation_2018}.

\section{Further enhancing KD with an optimal tree algorithm}\label{sec:evtrees}
\subsection{Greedy and optimal trees}
There is one part of this KD process that I have glossed over so far, that is the tree fitting algorithm itself. This is because the knowledge distillation literature is agnostic to the actual learner used to fit a distilled tree. Most papers (such as Dao et al.) tend to use a greedy algorithm. Greedy tree fitting algorithms like the Classification and Regression Tree (CART) \citep{breiman_classification_2017} are by far the most widely tree learners in real-world applications. These trees partition greedily, that is they start at the root node of the tree, find the split that best minimises the splitting criterion (e.g. mean squared error), then continue recursively partitioning, looking every time at just how to minimise the criterion at one specific node. This of course risks that a particular node might seem best at the start of the process but not actually be the ideal variable to split on. 

Optimal splitting approaches instead optimise over every split at the same time. There are two reasons this is not widely done in practice. The first is that fitting an optimal tree is an extremely computationally challenging problem, much harder than greedy splitting. In fact it is in the class of \textit{np}-hard problems meaning that it quickly becomes infeasible as the number of observations, features, or the size of the tree increases \citep{bertsimas_optimal_2017}. The other reason is that for applications that do not place a high value on interpretability (i.e. most machine learning applications \citep{rudin_stop_2019}) it is generally a much better use of computational resources to simply fit a large random forest (or some kind of other ensemble like a boosting model) rather than expending the resources to optimise a single tree \citep{breiman_random_2001}. Many greedy learners simply outperform one optimal one \citep{rudin_stop_2019}.

However, in our particular application, these drawbacks should give us less pause. The reason for this is that we care about interpretability of our models and so want to limit the complexity of an individual learner \citep{rudin_stop_2019}. This means we can only fit one tree, so we can't take advantage of an ensemble. We also generally want to constrain the complexity of our single decision tree to keep it interpretable, so it may be worth finding out whether the structure is simple enough to make an optimal tree approach practical. Under these constraints, an optimal tree can be a useful model with better fit of training data, and greater stability.

Recent improvements in algorithms and computing power have made optimal trees tractable for moderately-sized problems (in terms of number of observations, number of features, and tree size) \citep{bertsimas_optimal_2017, demirovic_murtree_2022}. There are a number of different technical approaches to doing this including dynamic programming, mixed integer programming and evolutionary approaches. Any of these approaches are possible for knowledge distillation. Again, distillation theory is agnostic to the actual learner used. However, for the purposes of this paper I will rely on a family of evolutionary tree learners, specifically the evtree algorithm \citep{grubinger_evtree_2014}. There are a few reasons for this. Firstly, there is a good implementation of evolutionary optimal trees (\textit{evtree}) already available in R so this will interface with \textit{grf} nicely. Secondly, it is easy to limit the number of iterations of the evolutionary algorithm to get a good-enough solution given a time constraint. In practice it may help to continue fitting until reaching a point of diminishing returns and then compare performance to a greedy distilled tree on distillation loss to see if it is worth taking the imperfectly optimised evolutionary tree or the greedy tree as the final student.

\subsection{How evolutionary trees work}
An evolutionary tree uses a process that is inspired by biological evolution to arrive at an optimal model. Here I will provide a brief, plain-English description of how the process works, but there is a lot of technical detail that will be omitted.\footnote{A more detailed explanation can be found in \url{https://cran.r-project.org/web/packages/evtree/vignettes/evtree.pdf}.} First, the learner generates some number of decision trees with completely random splits to form the initial population, then it begins iterating through an optimisation procedure. This procedure involves randomly applying one of a number of possible variation operations to each tree in the population 

\begin{itemize}
  \item Split -- Add a random split to a terminal node of the tree.
  \item Prune -- Remove two terminal nodes with the same parent, making the parent node a new terminal node.
  \item Major split rule mutation -- Change a random split rule. There is a 50\% chance of changing the split variable along with the split rule and a 50\% chance of just changing the split rule. These changes are fully random.
  \item Minor split rule mutation -- Leaves split variable unchanged and makes a minor perturbation to the splitting rule.
  \item Crossover -- randomly swaps two parts (subtrees) of a the selected tree and another random tree.
\end{itemize}

The trees are all then evaluated with an evaluation function. This evaluation function is generically a loss function penalised by a complexity function $$\hat{\theta} = \arg\min_{\theta \in \Theta} \text{loss}\{Y, f(X, \theta)\} + \text{comp}(\theta).
$$Here $\theta$ is the structure of the tree (splitting values, and variables for each non-terminal node) and $\Theta$ is the parameter-space within which we can find splits.

For regression applications \citet{grubinger_evtree_2014} suggests using a penalised MSE where $N$ is the number of datapoints, $M$ is the number of terminal nodes, and $\alpha$ is a complexity penalising hyperparameter,$$
\text{loss}(Y, f(X, \theta)) = N \log \text{MSE}(Y, f(X, \theta)) = N \log \left\{ \sum_{n=1}^{N} (Y_n - f(X \cdot n, \theta))^2 \right\},$$
$$\text{comp}(\theta) = \alpha \cdot 4 \cdot (M + 1) \cdot \log N.
$$

Once trees are evaluated it is necessary to choose the trees that will proceed to the next generation. While there are a number of possible approaches, \textit{evtree} uses deterministic crowding where each parent competes against its direct child (the one formed from it by the variation operator). Whichever tree performs better on the evaluation function proceeds to the next generation.

The process stops once the population converges on a solution. The algorithm is judged to have converged when the top performing 5\% of trees does not change for 100 iterations, but not before 1000 iterations \citep{grubinger_evtree_2014}. Once the algorithm converges, the best-performing tree is returned. If the algorithm does not converge in the time allowed (by default 10,000 iterations), the best performing tree is returned after reaching this limit.

\subsection{Limitations of evolutionary trees}
Given we want to optimise under strict model complexity constraints, there is a strong case for an optimal model and an evolutionary tree allows for optimal fitting with flexibility around fitting more complex tree structures. However, it still has its drawbacks. The two main ones are around picking the hyperparameters to control splitting and dealing with missing data.

The standard causal tree works well with missing data. It can decide which way to classify observations with missing values for a split variable meaning that there can be information in missingness that is taken into account in estimation. This is not the case for evolutionary trees which do not cope with missing data. Here missing data must be imputed in some way. This leads to a question then of whether it is better to fit a greedy model that can get information from missingness or an optimal model that cannot. In a dataset with lots of missing data, it may be an important choice, but fortunately distillation loss can proxy for performance on ground-truth to compare the two different approaches.

There is also a question of whether evolutionary optimisation is even worth it in high-dimensional datasets (i.e. one where $\Theta$ is large). It is possible to imagine that for a high-dimensional dataset, random permutation of a finite number of trees, over a finite number of generations may struggle to explore $\Theta$ effectively as there are simply too many such models. It may be worth simply exploring this space under the limitations of greediness. Again, distillation loss can be a guide here. If the evolutionary algorithm is performing poorly on ground-truth treatment effects, this should be clear from its poor performance on distillation loss compared to a greedy tree.

\section{Simulation study}\label{sec:sim}
\subsection{Set-up}
Unfortunately, it is difficult to quantify the extent to which a given tree gives an interpretable picture of the underlying CATE DGP. However, a good proxy for this might be the power of an interpretable tree fit under a set complexity limit to predict treatment effects on unseen data.

In this section we consider the performance of the DCT approach on synthetic data. Synthetic data is useful because it allows us to know the ground-truth causal effects. This section uses a set of DGPs is from the Atlantic Causal Inference Competition (ACIC) 2016 dataset which was designed to be a realistic but challenging set of DGPs to test the performance of causal machine learning models. There has already been a full write-up of this competition and the methods employed in it \citep{dorie_automated_2019}.

Note that at no point is any learner in this simulation given access to the ground-truth causal effects, the models are fit using a standard causal forest and then the predictions from that forest are used as the target of the student learner in the knowledge distillation. The ground-truth effects are simply used to evaluate performance.

The data generated for ACIC 2016 had a couple of desirable qualities. It was meant to be calibrated to real data so the data-generating process would be a somewhat realistic process. These data generating processes were also complex and nonlinear, meaning that this is not an easy a task for the learners that will stop us from being able to distinguish different approaches from each other. The data is also easily available. The version of the data used for this simulation is taken from the built-in data in the Causal Inference 360 Python package.\footnote{\url{https://raw.githubusercontent.com/BiomedSciAI/causallib/master/causallib/datasets/data/acic\_challenge\_2016}} Each DGP has 4802 rows and 58 covariates. The data will be randomly split 50-50 into a train and test set. Each DGP also has a varying amount of complexity in its CATE distribution and its confounding structure \citep{dorie_automated_2019}.

All individual trees here will be fit (or pruned) to a maximum depth of four --- there are two reasons to do this. Firstly, in order to maximise out-of-sample fit, individual trees need to be explicitly regularised in a way trees used a random forest do not \citep{breiman_random_2001}. This means it is worth limiting the size of the distilled trees in some way. This also means it is not fair to the trees pulled from the forest to take an unpruned tree and try to predict out to new data in a comparison with a depth-limited tree, so we will prune each extracted tree. The second point is that when interpreting a tree, we are unlikely to look much further than the first few layers of splits anyway. A very large tree can be uninterpretable due to its complexity \citep{rudin_stop_2019, lipton_mythos_2018}. Realistically, a trimmed tree is likely to be a more interpretable tree.

Several models are included in this comparison, including the greedy and optimal DCTs. In addition being pruned to depth 4, all trees also predict out using the tree structure as an adaptive kernel for doubly robust estimation akin to the DCT or the causal forest. All models use the same 10,000 tree nuisance models for doubly-robust estimation (and local centering if the model uses local centering). The models compared in this simulation are:
\begin{itemize}
    \item The full teacher forest with 10,000 trees.
    \item An optimal DCT with 400,000 generations and 800 trees considered each generation.
    \item A greedy DCT fit with CART.
    \item A tree selected from the ensemble via the \citet{wager_find_2018} method of picking a causal tree. This searches the ensemble for the tree that best minimises R-Loss \citep{nie_quasi-oracle_2020}, the objective used to fit a causal forest.
    \item The average of all the pruned trees in the ensemble.
    \item A small forest with 10 trees.
    \item A single non-distilled tree (a one tree causal forest, fit on the raw data and trimmed to depth 4). This is one way of testing the approach of \citet{giannarakis_towards_2022, rana_machine_2019} who fit a single undistilled causal tree on raw data.
\end{itemize}

This provides reasonable coverage of all the existing methods I am aware of that have been used for finding a single, representative tree \citep{rehill_2024_how)}. Wager's approach of course is pulled directly from the code provided on the \textit{grf} github \citep{wager_find_2018}. The random tree approach used by \citet{amann_effect_2023} is represented not by the pulling of a single tree, but by showing the distribution of predictions all trees (which is also the probability distribution for the random tree approach). The only other proposed approach I am aware of that is not tested here is from \citet{zhou_heterogeneous_2023} and \citet{zhou_targeting_2023} which use a tree distance approach from \citet{banerjee_identifying_2012} to find the 'most representative' tree in an ensemble. It is unclear how these papers construct a tree-distance metric for the causal forest as they do not provide code for their approach or details on implementation. The distribution of all trees and the mean of this distribution should hopefully provide a sense of how well this approach might work.

This section will mostly be concerned with how well the DCT works in predicting ground-truth causal effects. However, I will also see how well these approaches perform on an R-Loss metric \citep{nie_quasi-oracle_2020}. This is because R-Loss provides pseudo-outcomes that can be used as an objective in CATE learning problems, but these are obviously only estimates made with nuisance models. If R-Loss is a reliable proxy for ground-truth performance though, that allows us to be more confident in choosing the best single-tree approach when we are using real-world data, because the best approach will likely be the R-Loss minimising model. All tests are run on held-out data. The sample is for each iteration split in half, training takes place on one half while testing takes place on the other.

\subsection{ACIC simulations}
\textbf{Results for ground-truth}

\begin{figure}[!h]
    \centering
    \includegraphics[width=1\linewidth]{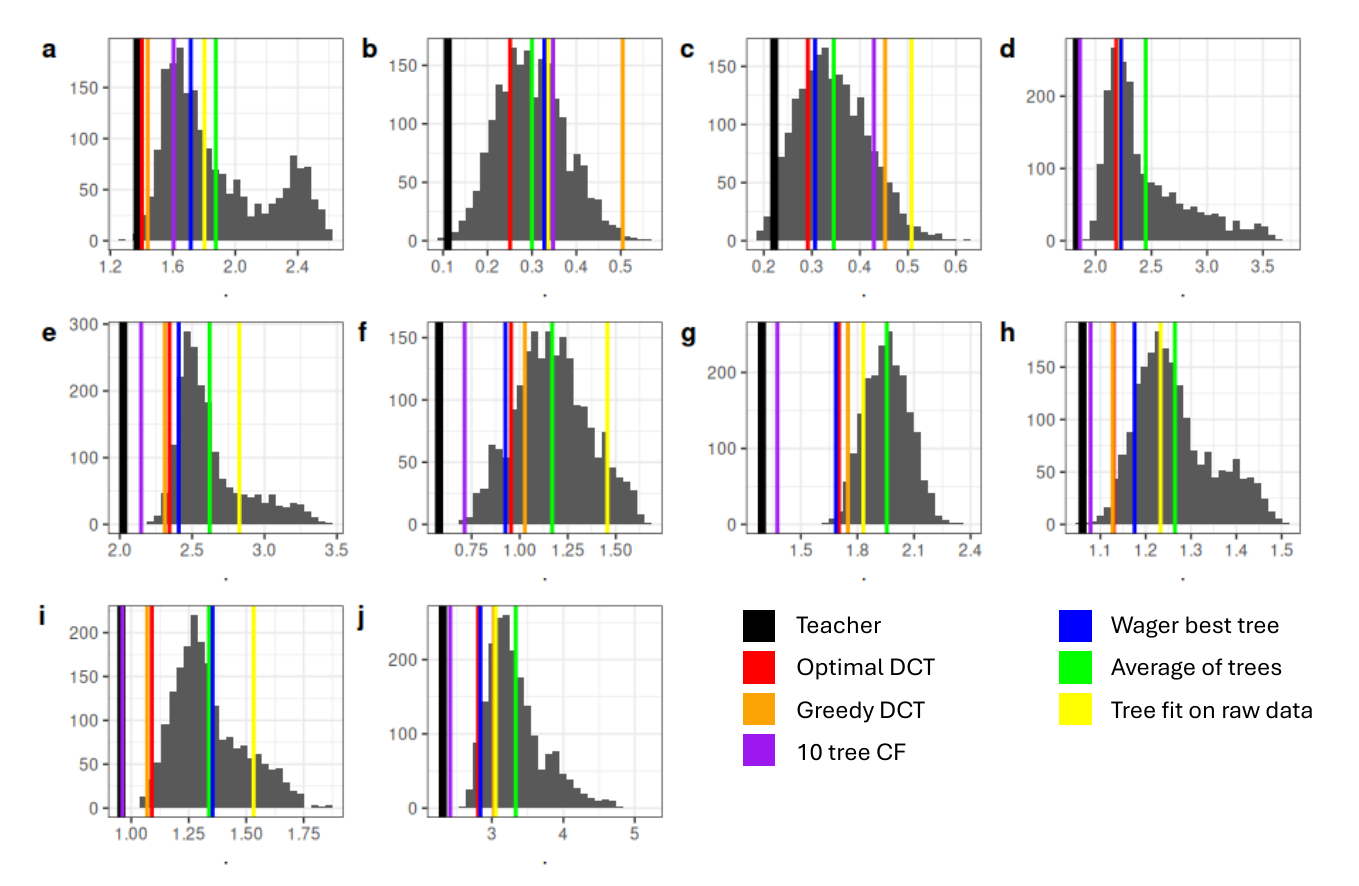}
    \caption{Ground truth mean absolute error results on original ACIC data. The grey histogram is predictions of the (pruned) individual trees in the ensemble while lines are the specific models for comparison.}
    \label{fig:gtnormal}
\end{figure}

We see in Figure \ref{fig:gtnormal} that on the original ACIC data, the the optimal distilled tree performs reasonably well. Generally, these slightly outperform the other single tree approaches including the mean tree outcome, the single tree fit on raw data, and the greedy distilled tree. However, they underperform the 10 tree forest and the teacher. Full results for all these analysis can be found in the Appendix.

\begin{figure}[!h]
    \centering
    \includegraphics[width=1\linewidth]{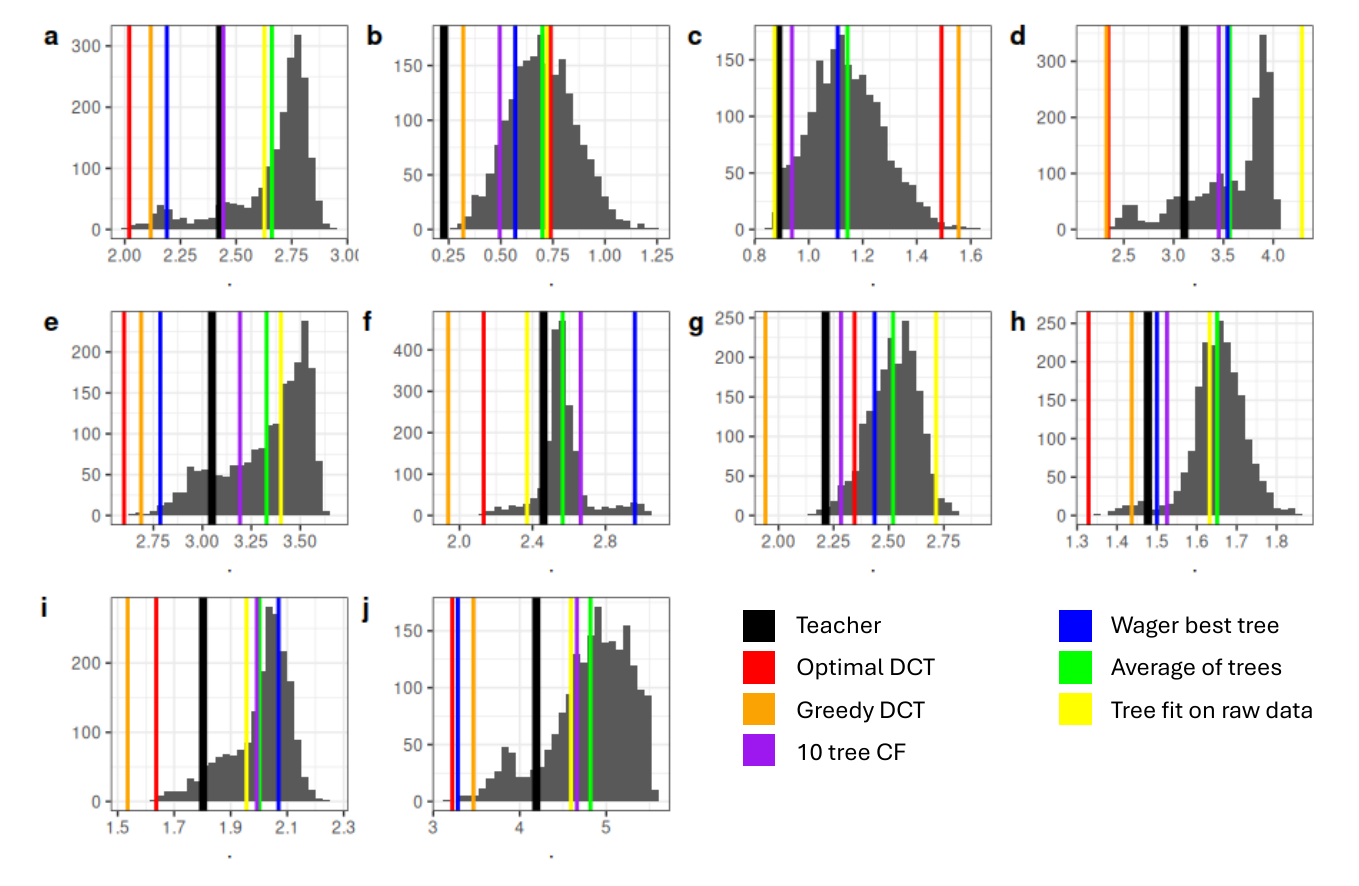}
    \caption{Ground truth mean absolute error results with noise and correlated features introduced into $X$. The grey histogram is predictions of the (pruned) individual trees in the ensemble while lines are the specific models for comparison.}
    \label{fig:gtnoisy}
\end{figure}

On the noisy data though, there is a marked difference as shown in Figure \ref{fig:gtnoisy}. Distillation approaches perform much better here. In most cases, the optimal and greedy trees are the leading approaches, beating out even the teacher ensemble. Interestingly, the R-Loss minimising tree seems to perform somewhat well here too (though a little worse than distillation). Here we see the vulnerability the causal forest has to correlated features as discussed in Section \ref{sec:correlation}. In this kind of environment, a well-chosen single tree, ideally from a distillation approach, might not actually be trading off interpretability for power at all. The tree that can deal better with correlated features might be better on both metrics.

\textbf{Results for R-Loss}

\begin{figure}[!h]
    \centering
    \includegraphics[width=1\linewidth]{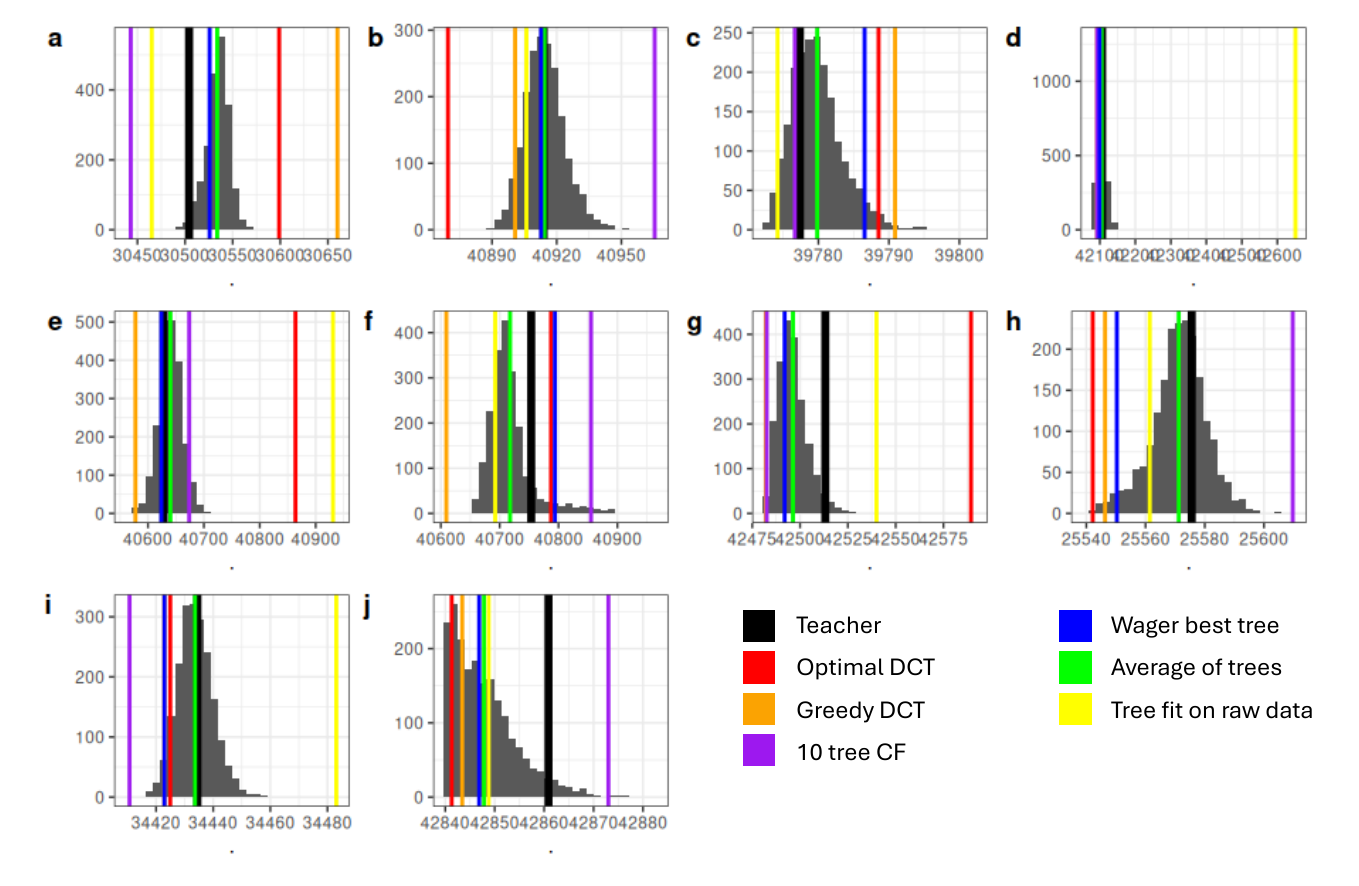}
    \caption{Simulation results on original ACIC data with R-Loss. The grey histogram is predictions of the (pruned) individual trees in the ensemble while lines are the specific models for comparison.}
    \label{fig:rnormal}
\end{figure}

\begin{figure}[!h]
    \centering
    \includegraphics[width=1\linewidth]{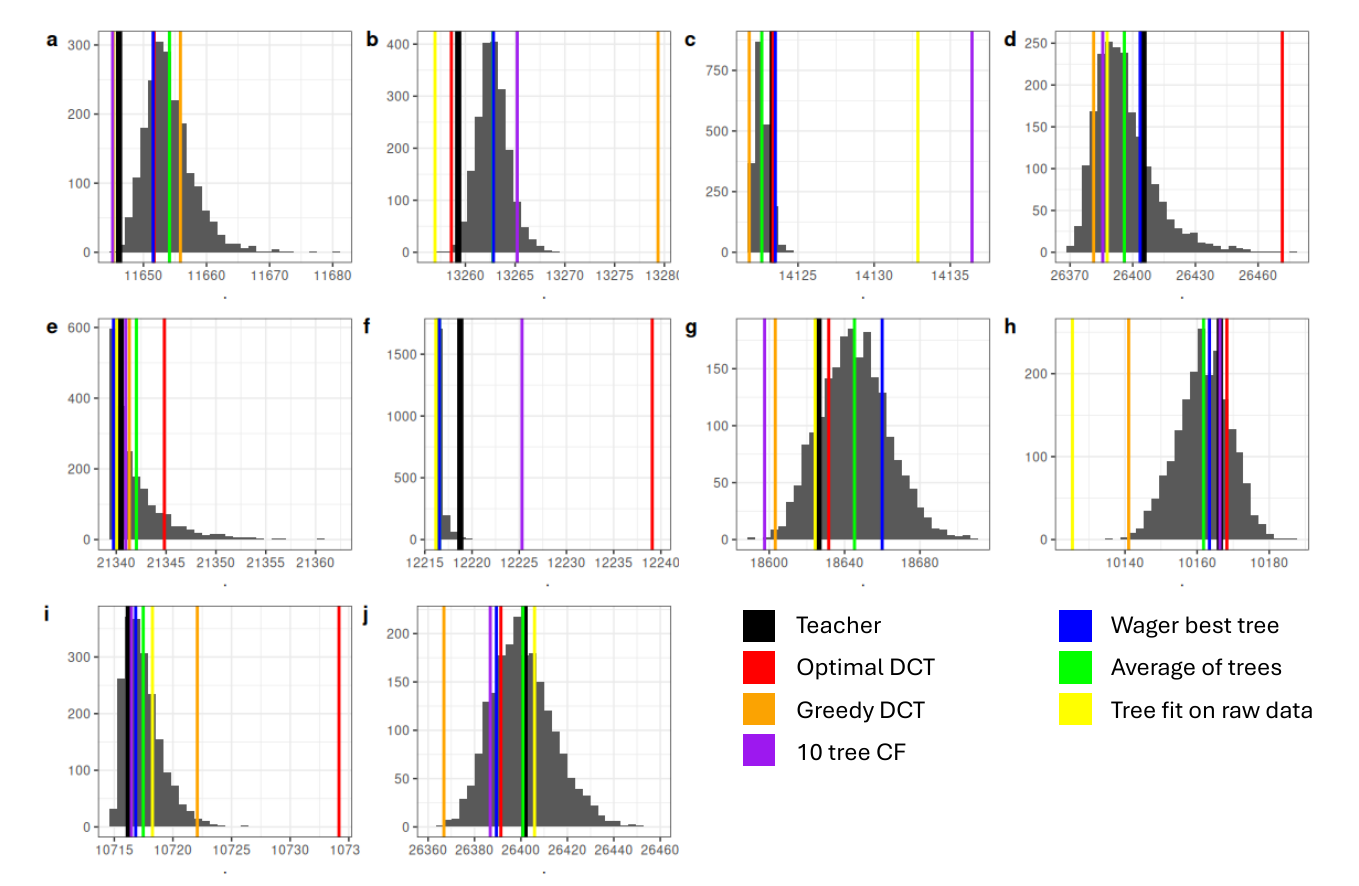}
    \caption{Simulation results with noise and correlated features introduced into $X$ with R-Loss. The grey histogram is predictions of the (pruned) individual trees in the ensemble while lines are the specific models for comparison.}
    \label{fig:rnoisy}
\end{figure}

Unfortunately, it seems that R-Loss is not a good metric to measure the performance of the distilled learner. The problems here seem to be noise in R-Loss and an upward bias in the Wager approach (which is tailored to minimising R-Loss). As can be seen in Figures \ref{fig:rnormal} and \ref{fig:rnoisy}, while the distilled trees perform well on ground-truth, the R-Loss metrics do not show this. The best tree by R-Loss metric over-performs here perhaps expectedly, but is generally worse on ground-truth outcomes. R-Loss is simply too noisy a metric to use to pick a tree extraction method. This suggests that there is no easy solution to picking a method that works well for a given dataset when using real-world data.

In this simulation then, the distilled, optimal approach generally outperforms other single tree approaches, particularly in high-dimensional datasets where there may be many irrelevant features or features correlated with each other. However, R-Loss does not provide an easy guide in identifying these kinds of scenarios. 

\FloatBarrier
\section{Application}\label{sec:application}
This section applies a DCT model to learn heterogeneous treatment effects using an example that has been used elsewhere to demonstrate heterogeneous treatment effect learners \citep{kunzel_metalearners_2019, green_machine_2023}. The data is taken from the \citet{broockman_durably_2016} field experiment on the effect of door-to-door canvassing on reducing transphobia in Miami voters. The experiment measured attitudes towards transgender people initially producing a score from 0 (negative) to 100 (positive). Then, the canvasser discussed trans issues with the participants in the treatment group or the benefits of recycling for the control group. Three days later, another person returned and again measured the individual's sentiment towards trans people to see if there was any effect.

\begin{figure}[ht!]
    \centering
    \includegraphics[width=0.75\linewidth]{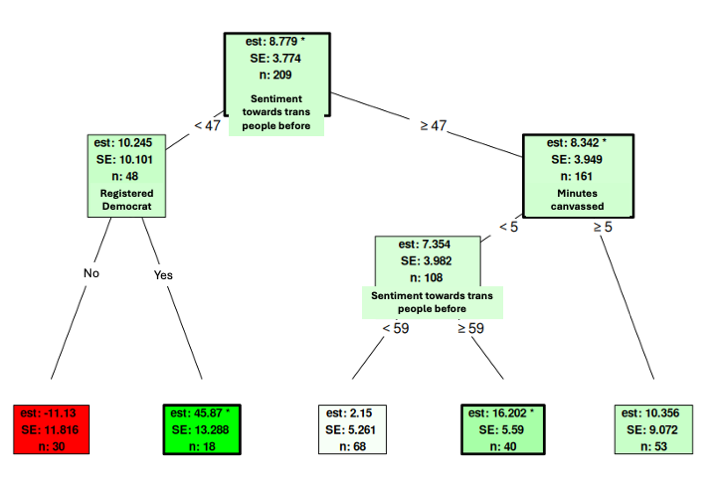}
    \caption{The DCT for the effect of a cash transfer on maths scores.}
    \label{fig:taydct}
\end{figure}

This application takes the data from this experiment with 419 people, and applies a 2000 tree causal forest taking sentiment towards trans people at the end of the conversation as the outcome and an array of 15 variables as predictors of heterogeneity (these cover party affiliation, race, a measure of social dominance orientation \citep{pratto_social_1994}, age, past voting behaviour, and time spent canvassing). For a full write-up of these variables see \citet{green_machine_2023} or \citet{broockman_durably_2016}. All non-continuous variables have been recoded as dummy variables. This was then distilled down into an optimal causal tree presented in Figure \ref{fig:taydct}.

The DCT presented here estimates out an effect in each node of the tree (in order to see both course and fine subgroup effects), colour-codes effects on a red-green scale and highlights statistically significant nodes (that is where we are 95\% confident the effect is not 0 based on bootstrapped standard errors) with a thicker border and asterisk on the effect estimate. The nodes display the estimate, standard error, number of individuals in that node (in the estimation sample) and the splitting variable for that node. The edges show the split values. The technical language of variable names (rather than labels) and binary encoded splits (e.g. "< 1" instead of "No" for dummy coded variables) have been changed for clarity.

This tree shows four nodes with a statistically significant effect. The first is the root node; which means there is a statistically significant ATE. This then splits on the basis of baseline sentiment towards trans people. Those with slightly higher sentiment have a somewhat lower but statistically significant effect here (perhaps because it is hard to raise scores when scores are already high). Finally of those who have lower starting sentiment at the base node, and were registered democrats, there was a substantial positive canvassing effect (though the sample size here was small). This is potentially due to this being a polarised issue and democrats who have low initial opinion may have partisan identities that have not yet been activated on this issue, but can be as a way to produce substantial gains \citep{kacholia_priming_2022}. Note though that the sample sizes here are relatively small.

This provides one way of possibly explaining heterogeneity. In reality there may be many other possible trees that perform almost as well. In settings where a tree is unstable (that is, nodes, particular those higher up the tree will change split variables or dramatically change split values in response to small perturbations in the data), it may be helpful to review a range of reasonably well-performing trees, for example by retrieving several high-performing trees from the evolutionary algorithm's final generation or by fitting the optimal algorithm on various sub-samples of the data. Then a researcher can try to understand a broader range of patterns in the data than might be clear from just a single tree. Of course, at a certain point, reviewing many trees hurts interpretability (and effectively makes this the same as interpreting an ensemble).

\section{Conclusion}
This paper proposed the Distilled Causal Tree as a method to extract out a single, interpretable tree which can be used to help understand a CATE distribution. The causal forest has the characteristics of a learner that should perform well as a teacher in knowledge distillation and so we can take advantage of this to create high performing, interpretable models. The simulations here show that the distillation approach fares well against other single tree approaches and performs particularly well in cases where there is noise and variables are correlated with each other.

A key finding here is that the distilled trees outperform the teacher in most of the noisy data simulations. This is notable because it means it is not enough to simply fit a causal forest on data and expect accurate predictions. This provides further impetus for the development of some kind of features selection method for the causal forest when working with high-dimensional datasets. There may also be benefits to experimenting with distillation in other ways (for example distilling a forest from another forest may improve the ability of the final forest to fit noisy data, akin to how the \citet{basu_iterative_2018} approach which is commonly applied to causal forests works \citep{rehill_2024_how)}).

The DCT approach of course still has its limitations. Fitting an optimal tree takes time and is computationally intensive. Furthermore, it is not necessarily clear to what extent the approach benefits from hyperparameter tuning. Hyperparameter tuning here will be even more computationally intensive because of the need to iterate through computationally costly models and this kind of tuning is likely outside the expertise of applied researchers using these methods (this is one reason why the grf library is so useful, being based on a method requiring relatively little tuning and having automatic tuning features built in). In fact, even though this paper used predictive performance of learners on causal effects as a proxy for the informativeness of a causal machine learning model, it is worth noting that the ultimate goal here is interpretability, not accuracy. There may be interpretability concerns (for example keeping trees to a low depth) that drive hyperparameter selection regardless of what the performance effect of these choices are.

The approach provided in this paper is a workable set of tools for distilling a causal tree, but it is easy to imagine how these could be improved on. For example, it is easy to imagine that performance of the optimal learner may be improved by having at least some of the trees in consideration initialised with a greedy fit. This is akin for example to how k-mean clustering is often used to initialise a Gaussian mixture model, use a cheap learner first then fine-tuning a more expensive learner. One issue with this could be models becoming stuck in local maxima around the greedy initialisations, but this could easily be solved with standard (random) initialisation for some trees.

In addition, while this paper limited itself to discussion of the causal forest model, this is not the only heterogeneous treatment effect learner that exists. Some do not naturally have the features that aid distillation, for example, meta-learners like X-Learner \citep{kunzel_metalearners_2019} lack the asymptotic unbiasedness of the causal forest which we might expect would make these models worse teachers all else held equal. In practice though, the greater flexibility of other approaches may be worth trading away some theoretical performance. For example, in uplift modelling, it is common to use meta-learners along with powerful base models like deep nets to get better performance despite worse theoretical guarantees \citep{zhang_unified_2022}. Distillation may still make sense here as a way of making sense of these models even if we would not expect the same accuracy from the student.

\bibliographystyle{agsm}  
\bibliography{references}

\section*{Appendix --- Detailed simulation results}

\begin{table}[!htbp] \centering 
  \caption{Simulation results for ground-truth effects with the regular DGPs.} 
  \label{tab:gt-simple} 
\begin{tabular}{@{\extracolsep{5pt}} cccccccc} 
\\[-1.8ex]\hline 
\hline \\[-1.8ex] 
 & Teacher & Optimal DCT & Greedy DCT & Wager best tree & 10 tree forest & Mean tree & Basic causal tree \\ 
\hline \\[-1.8ex] 
DGP a & 1.374 & 1.402 & 1.44 & 1.715 & 1.605 & 1.876 & 1.802 \\ 
DGP b & 0.110 & 0.251 & 0.504 & 0.328 & 0.347 & 0.300 & 0.336 \\ 
DGP c & 0.222 & 0.292 & 0.453 & 0.307 & 0.429 & 0.346 & 0.508 \\ 
DGP d & 1.832 & 2.185 & 2.226 & 2.226 & 1.858 & 2.449 & 2.216 \\ 
DGP e & 2.024 & 2.343 & 2.312 & 2.408 & 2.148 & 2.621 & 2.826 \\ 
DGP f & 0.581 & 0.955 & 1.027 & 0.926 & 0.714 & 1.169 & 1.456 \\ 
DGP g & 1.293 & 1.702 & 1.750 & 1.686 & 1.375 & 1.956 & 1.831 \\ 
DGP h & 1.061 & 1.129 & 1.127 & 1.175 & 1.078 & 1.265 & 1.233 \\ 
DGP i & 0.957 & 1.090 & 1.069 & 1.354 & 0.961 & 1.338 & 1.532 \\ 
DGP j & 2.314 & 2.808 & 3.030 & 2.840 & 2.417 & 3.334 & 3.053 \\ 
\hline \\[-1.8ex] 
\end{tabular} 
\end{table}

\begin{table}[!htbp] \centering 
  \caption{Simulation results for ground-truth effects with the noisy DGPs.}
  \label{tab:gt-noisy} 
\begin{tabular}{@{\extracolsep{5pt}} cccccccc} 
\\[-1.8ex]\hline 
\hline \\[-1.8ex] 
 & Teacher & Optimal DCT & Greedy DCT & Wager best tree & 10 tree forest & Mean tree & Basic causal tree \\ 
\hline \\[-1.8ex] 
DGP a & 2.428 & 2.020 & 2.116 & 2.189 & 2.442 & 2.661 & 2.627 \\ 
DGP b & 0.227 & 0.739 & 0.320 & 0.569 & 0.495 & 0.699 & 0.721 \\ 
DGP c & 0.887 & 1.491 & 1.555 & 1.107 & 0.938 & 1.144 & 0.874 \\ 
DGP d & 3.108 & 2.344 & 2.327 & 3.541 & 3.455 & 3.564 & 4.289 \\ 
DGP e & 3.049 & 2.600 & 2.687 & 2.785 & 3.192 & 3.328 & 3.401 \\ 
DGP f & 2.461 & 2.133 & 1.938 & 2.961 & 2.665 & 2.566 & 2.370 \\ 
DGP g & 2.214 & 2.345 & 1.941 & 2.436 & 2.283 & 2.520 & 2.715 \\ 
DGP h & 1.477 & 1.329 & 1.437 & 1.500 & 1.525 & 1.651 & 1.632 \\ 
DGP i & 1.802 & 1.636 & 1.535 & 2.069 & 1.993 & 2.002 & 1.956 \\ 
DGP j & 4.191 & 3.220 & 3.464 & 3.284 & 4.661 & 4.817 & 4.592 \\ 
\hline \\[-1.8ex] 
\end{tabular} 
\end{table}

\begin{table}[!htbp] \centering 
  \caption{Simulation results for R-Loss with the regular DGPs.}
  \label{tab:rl-simple} 
\begin{tabular}{@{\extracolsep{5pt}} cccccccc} 
\\[-1.8ex]\hline 
\hline \\[-1.8ex] 
 & Teacher & Optimal DCT & Greedy DCT & Wager best tree & 10 tree forest & Mean tree & Basic causal tree \\ 
\hline \\[-1.8ex] 
DGP a & 11646.137 & 11651.698 & 11655.863 & 11651.545 & 11645.124 & 11654.129 & 11645.280 \\ 
DGP b & 13259.279 & 13258.593 & 13279.365 & 13262.819 & 13265.215 & 13262.772 & 13256.949 \\ 
DGP c & 14123.339 & 14123.353 & 14121.797 & 14123.511 & 14136.433 & 14122.623 & 14132.877 \\ 
DGP d & 26405.163 & 26471.603 & 26381.281 & 26403.551 & 26385.665 & 26395.997 & 26387.767 \\ 
DGP e & 21340.464 & 21344.826 & 21341.308 & 21339.725 & 21340.946 & 21342.022 & 21340.031 \\ 
DGP f & 12218.758 & 12239.083 & 12216.358 & 12216.556 & 12225.294 & 12216.542 & 12216.167 \\ 
DGP g & 18626.158 & 18631.588 & 18603.224 & 18659.947 & 18597.553 & 18645.250 & 18624.409 \\ 
DGP h & 10166.303 & 10168.258 & 10141.020 & 10163.424 & 10166.333 & 10161.792 & 10125.421 \\ 
DGP i & 10716.236 & 10734.166 & 10722.082 & 10716.826 & 10716.455 & 10717.470 & 10718.255 \\ 
DGP j & 26401.563 & 26391.360 & 26366.827 & 26389.433 & 26386.789 & 26400.862 & 26405.860 \\ 
\hline \\[-1.8ex] 
\end{tabular} 
\end{table}

\begin{table}[!htbp] \centering 
  \caption{Simulation results for R-Loss with the noisy DGPs.} 
  \label{tab:rl-noisy} 
\begin{tabular}{@{\extracolsep{5pt}} cccccccc} 
\\[-1.8ex]\hline 
\hline \\[-1.8ex] 
 & Teacher & Optimal DCT & Greedy DCT & Wager best tree & 10 tree forest & Mean tree & Basic causal tree \\ 
\hline \\[-1.8ex] 
DGP a & 30504.245 & 30598.794 & 30660.074 & 30525.787 & 30442.963 & 30533.739 & 30465.003 \\ 
DGP b & 40913.948 & 40869.578 & 40900.721 & 40912.798 & 40965.446 & 40914.288 & 40905.955 \\ 
DGP c & 39777.376 & 39788.537 & 39790.877 & 39786.552 & 39776.658 & 39779.829 & 39774.208 \\ 
DGP d & 42106.357 & 42094.609 & 42092.382 & 42099.345 & 42092.352 & 42105.077 & 42650.994 \\ 
DGP e & 40628.047 & 40863.878 & 40577.611 & 40624.06 & 40673.799 & 40640.138 & 40931.0290 \\ 
DGP f & 40753.568 & 40787.246 & 40609.357 & 40793.818 & 40855.205 & 40717.663 & 40692.361 \\ 
DGP g & 42513.202 & 42589.423 & 42482.053 & 42491.943 & 42482.548 & 42496.297 & 42539.981 \\ 
DGP h & 25575.603 & 25542.161 & 25546.339 & 25550.315 & 25609.713 & 25571.227 & 25561.430 \\ 
DGP i & 34434.472 & 34424.961 & 34422.897 & 34422.929 & 34410.722 & 34433.655 & 34483.081 \\ 
DGP j & 42860.886 & 42841.295 & 42843.426 & 42846.858 & 42873.005 & 42847.834 & 42848.775 \\ 
\hline \\[-1.8ex] 
\end{tabular} 
\end{table} 

\end{document}